\documentclass[sigconf]{acmart}
\usepackage[T1]{fontenc}

\usepackage{color}

\newcommand{\enquote}[1]{``#1''}

\usepackage{xspace}

\usepackage[inline]{enumitem}

\usepackage{hyperref}
\usepackage[nameinlink]{cleveref}

\newcommand{\tablebullet}{{\LARGE \textbullet}}
\newcommand{\tablebulletNegative}{{\large \textopenbullet}}

\newcommand{\rot}[1]{\rotatebox{90}{#1}}

\newcommand{\conditions}{conditions\xspace}

\acmYear{2024}\copyrightyear{2024}
\setcopyright{rightsretained}
\acmConference[ACM FAccT '24]{ACM Conference on Fairness, Accountability, and Transparency}{June 3--6, 2024}{Rio de Janeiro, Brazil}
\acmBooktitle{ACM Conference on Fairness, Accountability, and Transparency (ACM FAccT '24), June 3--6, 2024, Rio de Janeiro, Brazil}
\acmDOI{10.1145/3630106.3659051}
\acmISBN{979-8-4007-0450-5/24/06}



\begin{document}
\begin{abstract}
    Human oversight is currently discussed as a potential safeguard to counter some of the negative aspects of high-risk AI applications.
    This prompts a critical examination of the role and conditions necessary for what is prominently termed \emph{effective} or \emph{meaningful} human oversight of these systems.
    This paper investigates effective human oversight by synthesizing insights from psychological, legal, philosophical, and technical domains. 
    Based on the claim that the main objective of human oversight is risk mitigation, we propose a viable understanding of effectiveness in human oversight: for human oversight to be effective, the oversight person has to have (a) sufficient causal power with regard to the system and its effects, (b) suitable epistemic access to relevant aspects of the situation, (c) self-control, and (d) fitting intentions for their role. Furthermore, we argue that this is equivalent to saying that an oversight person is effective if and only if they are morally responsible and have fitting intentions.
    Against this backdrop, we suggest facilitators and inhibitors of effectiveness in human oversight when striving for practical applicability. We discuss factors in three domains, namely, the technical design of the system, individual factors of oversight persons, and the environmental circumstances in which they operate. Finally, this paper scrutinizes the upcoming AI Act of the European Union -- in particular Article 14 on Human Oversight -- as an exemplary regulatory framework in which we  study the practicality of our understanding of effective human oversight. By analyzing the provisions and implications of the European AI Act proposal, we pinpoint how far that proposal aligns with our analyses regarding effective human oversight as well as how it might get enriched by our conceptual understanding of effectiveness in human oversight. 
    
    \end{abstract}
\title[On the Quest for Effectiveness in Human Oversight: Interdisciplinary Perspectives]{On the Quest for Effectiveness in Human Oversight:\\Interdisciplinary Perspectives}

\author{Sarah Sterz}
\affiliation{  
   \institution{Dependable Systems and Software, Saarland University}
   \city{Saarland Informatics Campus, Saarbrücken}
   \country{Germany}}
\email{sterz@depend.uni-saarland.de}

\author{Kevin Baum}
\affiliation{
   \institution{Neuro-Mechanistic Modeling, German Research Center for Artificial Intelligence (DFKI)}
   \city{Saarbrücken}
   \country{Germany}}
\email{kevin.baum@dfki.de}

\author{Sebastian Biewer}
\affiliation{
   \institution{Dependable Systems and Software, Saarland University}
   \city{Saarland Informatics Campus, Saarbrücken}
   \country{Germany}}
\email{biewer@depend.uni-saarland.de}

\author{Holger Hermanns}
\affiliation{
   \institution{Dependable Systems and Software, Saarland University}
   \city{Saarland Informatics Campus, Saarbrücken}
   \country{Germany}}
\email{hermanns@cs.uni-saarland.de}

\author{Anne Lauber-R\"onsberg}
\affiliation{
   \institution{IRGET, Faculty of Humanities and Social Science, TU Dresden}
   \city{Dresden}
   \country{Germany}}
\email{anne.lauber-roensberg@tu-dresden.de}

\author{Philip Meinel}
\affiliation{
   \institution{IRGET, Faculty of Humanities and Social Science, TU Dresden}
   \city{Dresden}
   \country{Germany}}
\email{philip.meinel@tu-dresden.de}

\author{Markus Langer}
\affiliation{
   \institution{Department of Psychology,\\ University of Freiburg}
   \city{Freiburg}
   \country{Germany}}
\email{markus.langer@psychologie.uni-freiburg.de}

\renewcommand{\shortauthors}{S. Sterz, K. Baum, S. Biewer, H. Hermanns, A. Lauber-Rönsberg, P. Meinel, and M. Langer}

\thanks{We want to thank Sven Hetmank and Franz Lehr for their contributions to previous work that this paper is based on. Also, we want to thank three anonymous reviewers for their valuable feedback.}

\maketitle   

\section{Introduction}

Effective or meaningful\footnote{Henceforth, we focus exclusively on the term \enquote{effective} and omit the term \enquote{meaningful}. Given their synonymous use in the context of human oversight, any statements regarding effectiveness also extend to meaningfulness. Furthermore, we are only interested in \emph{human} oversight, so we occasionally drop the \enquote{human} and talk about \enquote{oversight} only for the sake of simpler wording. In response to a very insightful reviewer comment, we have also avoided the phrase \enquote{human overseer} because of its problematic connotation in US-American contexts.
} human oversight is a notorious centerpiece of ethical guidelines \cite{Jobin_2019} and global legislation \cite{ai-act-ParliamentCOR, green-flaws} intended to govern the deployment of AI-based systems, particularly in high-risk contexts. The regulatory significance of human involvement in automated decision-making is unmistakable, as evidenced by a range of legislation \cite{green-flaws, enqvist2023human}, including the overarching principles embedded in the General Data Protection Regulation of the EU \cite{GDPR} and more targeted legislation, such as the European AI Act \cite{ai-act-ParliamentCOR}, which mandates human oversight for high-risk AI applications. Research has made strides in defining and assigning responsibility for human oversight \cite{enqvist2023human}, or placing it within a broader risk management framework \cite{green-flaws}. Human oversight has also been subject to criticism, with concerns being raised about the efficacy of oversight in practice and its potential role as a mere vehicle to legitimize imperfect AI systems \cite{green-flaws, Zerilli_2019}. 

Effective oversight has been discussed by many \cite{green-flaws, enqvist2023human, laux2023, Methnani2021}; however, a question that is crucial for this discussion has not been sufficiently addressed: \emph{When is human oversight effective?}
The pressing need for conceptual clarity in this regard is evident, particularly given that the European Union's AI Act is due to be adopted in 2024. Nevertheless, the definition and conditions for achieving \emph{effective} human involvement remain underexplored, highlighting a critical gap in the current state of the art of AI governance. 

The following core contributions are made by this paper, and as such provide a structural frame for it:
\begin{itemize}
    \item  We suggest a viable understanding of effectiveness in the context of human oversight based on the idea that the central objective of oversight is risk mitigation. Namely, we propose that an oversight person is effective if and only if they have (a) sufficient causal power with regards to the system and its effects, (b) suitable epistemic access to relevant aspects of the situation, (c) self-control over their own actions, and (d) fitting intentions for their role.  We point out that (a)--(c) are jointly sufficient for moral responsibility.    
    
    So, phrased boldly, we could say that
    
    \par
    \begin{center}
        {Effectiveness $=$ Moral Responsibility $+$ Fitting Intentions.}
    \end{center}
       
    \item We exemplify facilitators and inhibitors of effectiveness in human oversight in three categories: the technical design of the system, the individual factors of the human who is in charge of oversight, and the environmental circumstances in which they operate. 
    
    \item As a litmus test, we look at the details of the European AI Act relative to our proposed conceptualization. We identify possible benefits and shortcomings of the AI Act.  In particular, we argue that our conceptualization provides both a more general \emph{and} a more practically useful conceptualization of effective human oversight compared to the respective stipulations in the AI Act.
\end{itemize}

\section{Related Work}

Researchers have spelled out who will be in charge of (ensuring) human oversight and have analyzed how human oversight fits into a more global risk management concept (as, e.g., in the AI Act~\cite{ai-act-ParliamentCOR}) together with other approaches on accurate and robust systems, data-privacy adhering systems, and transparent systems \cite{enqvist2023human, laux2023}. The concept of human oversight, however, also faces substantial criticism, as empirical evidence suggests that humans may be ineffective in supervising AI-based systems \cite{green-flaws}. This has prompted a call for a more institutional oversight approach, emphasizing the need for empirical evidence regarding humans' capability to oversee AI systems prior to enforcing the oversight~\cite{green-flaws}. This section will explore the literature on effective oversight from different disciplines in more detail.

\paragraph{Effective human oversight in legislation and guidelines} 
Oversight is an essential aspect of various ethical guidelines and legislation on the use of AI-based systems in high-risk contexts. 
After the analysis of 41 policy documents around the globe, Green \cite{green-flaws} observes a call for limiting solely automated decisions (e.g., Article 22 of the European General Data Protection Regulation \cite{GDPR}), and an emphasis on the need for human discretion in AI-based decision-making (e.g., Canadian Directive on Automated Decision-Making \cite{Canada-Directive-Automated-Decision-Making}). 
The requirement for effective oversight is particularly highlighted in the European AI Act. At the time of writing, the AI Act has not yet been formally adopted. However, a final version was approved by the EU Parliament in March 2024 \cite{ai-act-ParliamentFINAL} and amended as part of the so-called corrigendum procedure \cite{ai-act-ParliamentCOR}. We will therefore use this amended version but will refer to earlier versions when relevant.

\paragraph{Human-centric AI} The focus on human oversight seems to stem from the background of ensuring human-centric development and deployment of AI-based systems. One of the most prominent concerns regarding the wide-range deployment of AI-based systems seems to be \enquote{that technological development and rationalized efficiency will take place at the cost of human agency and safety or rights} \cite{enqvist2023human}. An obvious way to counter such a techno-centric deployment of AI would be a human-centric development and deployment of AI-based systems. As Enqvist \cite{enqvist2023human} notes: \enquote{Human centricity, as it has come to be (broadly) understood, does not only reflect that human needs are to be met by new technologies, but also incorporates the aim to safeguard individual rights and increase human well-being.} 
Indeed, in computer science, a popular area of research is concerned with \emph{human-centered AI} (HCAI) where efficient collaboration between humans and AI systems is investigated \cite{Bansal_2019,Bu_inca_2021, Ehsan_2022,Lai_2023, Schoeffer_2023, tahaei2023human}. 

\paragraph{Objectives of effective human oversight}  

One of the foundational assumptions for the effectiveness of oversight measures is that human oversight should not be an end in itself \cite{laux2023}, but that the involvement of humans in contexts where AI-based systems affect high-risk decision-making can make things better in some regard. For example, one hope is that humans may be better than AI-based systems at incorporating ethical considerations and social norms into decision-making contexts \cite{enqvist2023human}. Another hope is that humans might be better at judging single or unusual cases, for instance, in decisions that would affect individual human beings \cite{Longoni_2019, baum2022responsibility}. Additionally, associating the term \enquote{effective} with human oversight shows the hope that a human can effectively contribute to a joint decision situation together with an AI-based system, thus enabling decisions better than those of either the human or the system alone \cite{Bansal_2021, Bu_inca_2021, Schoeffer_2023}. Other hopes include increased safety \cite{Koulu2020ProceduralizingCA}, improved accuracy \cite{jones2017}, or better trustworthiness and more actual trust \cite{Methnani2021}. The most important objective of human oversight recently seems to be the mitigation of risks to fundamental rights, both directly (as one reading of the AI Act \cite{ai-act-ParliamentCOR} suggests), or indirectly via, e.g., the goal to improve safety or the avoidance of ethically and socially undesirable outcomes \cite{enqvist2023human}.

\paragraph{Criticism of human oversight} However, these foundational assumptions have received substantial criticism, calling into question the usefulness of human oversight \cite{green-flaws, Zerilli_2019}. Critics (in particular \cite{green-flaws}) point to a lack of empirical evidence showing that we, 
as of now, can reliably integrate human and system abilities in a way that leads to better decisions compared to decisions that every single entity would produce in isolation \cite{Bartlett_2020, Bansal_2021, Rieger_2022}. For instance, there seem to be substantial challenges with respect to achieving adequate trust in AI-based systems \cite{Bansal_2019,lee2004, hoff2015trust, Zhang_2020}. Humans may overly rely on the outputs produced by automated systems, leading to situations where they do not detect erroneous or unfair outputs \cite{Parasuraman_2010}. In other situations, humans may have too little trust in AI-based systems or too high confidence in their own abilities, which leads to people overriding actually accurate or fair system outputs \cite{Green_2019}. Especially in high-reliability contexts, automation bias (cf. \Cref{sec:automationbias}) is a likely-to-observe phenomenon where people tend to use the outcome of an automated decision entity “as a heuristic replacement for vigilant information seeking and processing” (\cite{Mosier_1996} p. 205). This makes it particularly challenging for people to detect and adequately address system failures or erroneous outputs. Whereas research has identified several avenues to improve human abilities to oversee systems (e.g., training of oversight persons \cite{Bahner_2008} or 
augmenting system outputs with confidence scores~\cite{Bahner_2008, Bansal_2019}), the positive effects associated with these avenues seem to be inconsistent \cite{green-flaws}. The criticism regarding human oversight could be expressed in an even more foundational way: The driving momentum behind many AI-based systems is the intention of producing decisions and actions that are, in some regard, better than what human experts could achieve \cite{green-flaws, Zerilli_2019}. Ironically, the very human beings whose imperfections are meant to be overcome by such an AI-based system end up in the role of overseers of the system~\cite{Zerilli_2019}. This is, for instance, a typical situation in certain areas of AI-based medical diagnostics 
\cite{HAGGENMULLER2021202, Nishida_2022}.

\paragraph{Positive effects of human oversight} However, it is also possible to draw more positive conclusions from the literature. 
There are relevant examples where human operators successfully identified errors made by automated systems~\cite{McBride_2013,De_Arteaga_2020}. In addition, research has enhanced our understanding of factors that affect human abilities to detect errors made by automated systems~\cite{De_Arteaga_2020,McBride_2013,Lai_2023,Schoeffer_2023}. Furthermore, ongoing research contributes to successfully unveiling design options to improve human-system collaboration \cite{Lai_2023,Vasconcelos_2023}. For example, recent work proposes to combine the strengths of machine and human decision-making by using so-called \emph{conformal predictions} for multi-label classifiers. There, the machine proposes a set of outputs (instead of a single output), which is guaranteed to contain the correct output with a probability of at least some pre-defined value~\cite{DBLP:conf/icml/StraitouriWOR23}. If this set contains only a single output, confidence is considered high enough to let the decision happen fully autonomously. If, instead, it contains more than one output, the choice is delegated to a human. It was shown that for certain tasks, the human-system decision accuracy indeed is higher when the human choice is restricted to the outputs that are in the proposed set~\cite{DBLP:journals/corr/abs-2306-03928}.

Overall, research reports mixed results regarding human oversight. However, the fact that effective human oversight is a popular requirement in legislation emphasizes the necessity of obtaining clarity on the very concept of effective human oversight. This clarity is crucial to better devise powerful laws and guidelines and for ensuring the successful and safe implementation of AI in high-risk contexts. 
Thus, in this paper, we aim to derive high-level \conditions for effectiveness in human oversight.

\paragraph{Moral Responsibility}
While the approach to effectiveness in human oversight that we suggest in this paper is the original result of collective insights of an interdisciplinary team of researchers from psychology, law, philosophy, and computer science, we also draw inspiration from several high-level conceptions of moral responsibility, as, e.g., in \cite{sep-moral-responsibility, sep-computing-responsibility, Fischer1998-FISRAC-3}. Most conceptions of moral responsibility have a focus on three aspects in common: a causal connection between the agent and the event for which the agent might be responsible, the agent having sufficient knowledge of their decision situation, and a certain autonomy of the agent \cite{sep-moral-responsibility, sep-computing-responsibility, Fischer1998-FISRAC-3}. This high-level conception is used as a rough guardrail in this paper. 

\section{What is Effective Human Oversight?}\label{sec:effectiveness}

We understand human oversight to be the supervision of a system by at least one natural person, typically with the authority to influence its operations or effects.\footnote{While this understanding of human oversight will serve as our working definition, it is not the only possible candidate (cf.\ \Cref{sec:limitations}). For further analysis of the concept of human oversight, see also \cite{laux2023}.} This influence operates at various levels throughout the system's lifecycle, including intervention during execution, reversal of faulty decisions, or adjustments of parameters to improve results \cite{Methnani2021}. Although the oversight person may be assigned additional responsibilities, such as monitoring system performance or making engineering decisions, these tasks are not themselves part of human oversight.
Even when having fixated a meaning for \enquote{human oversight}, it is not clear what \enquote{effective} means in this regard. 

\subsection{An objective-first approach to the effectiveness of human oversight}

We want oversight to be useful. Therefore, it makes sense to define the effectiveness of human oversight in terms of how well it achieves its objectives. As we have discussed above, risk mitigation can be taken to be one of the main objectives of human oversight. This is the objective we will be focusing on. Additionally, effective human oversight arguably may also yield other positive effects, such as improved incorporation of ethical considerations \cite{enqvist2023human}, or enhanced judgment in cases impacting individual well-being \cite{Longoni_2019, baum2022responsibility}.

It is important to note that the understanding of effectiveness we are about to suggest is not the only viable one. While offering a structured framework, it is just one among potentially numerous conceptualizations.  
One way to change our conceptualization of effectiveness in human oversight would be to put objectives other than risk mitigation into focus, such as improved accuracy, trustworthiness, actual trust, human autonomy, accountability and responsibility, or liability. Discussing them is beyond the scope of this paper; however, \emph{prima facie}, there appears to be no inherent conflict between our proposal and these objectives. Furthermore, we believe that our conceptualization will be helpful beyond risk mitigation since it will also be able to address other objectives of human oversight, including (moral) responsibility, as we will demonstrate in \Cref{sec:responsibility}.

\begin{figure*}
    \centering
    \includegraphics[width=0.85\linewidth]{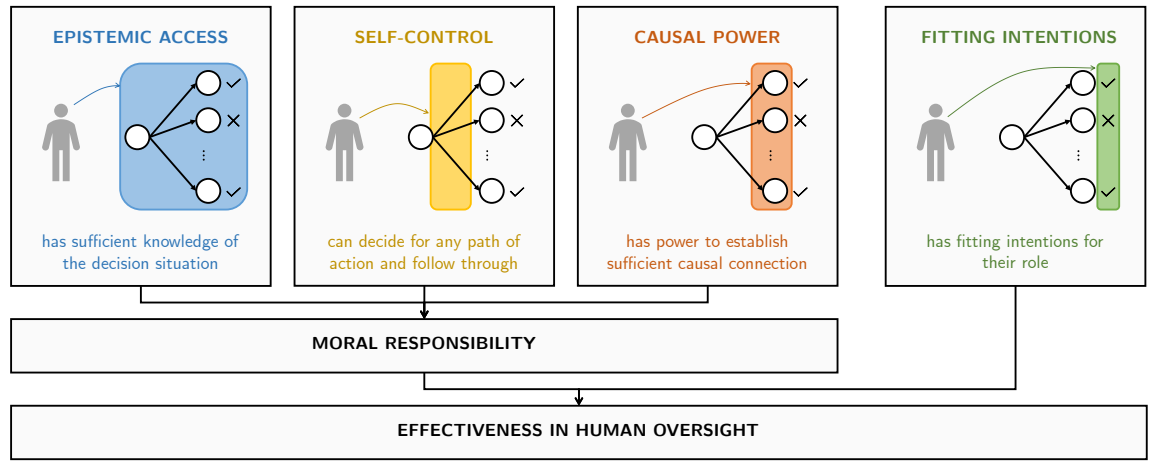}
    \Description{A diagram that shows the four conditions of effectiveness in human oversight: the boxes for epistemic access, self-control, causal power have outgoing arrows that jointly point to another box for moral responsibility; the box for moral responsibility and the box for fitting intentions have outgoing arrows that jointly point to another box for effectiveness in human oversight.}
    \caption{The four \conditions of effectiveness in human oversight and their relation to moral responsibility. The person in the diagram depicts the human in charge of oversight and the graph a schematic depiction of their current decision situation: the node on the left is their current position, the edges are the actions that they could perform, and the nodes on the right are the consequences of their actions.} 
    \label{fig:effFlowDiagram}
\end{figure*}

In line with the proposition that the main objective of effective human oversight is to mitigate risk, the main question is: \emph{When does human oversight facilitate the mitigation of risks?} We propose this is the case if the oversight person meets the following four key \conditions: causal power, epistemic access, self-control, and fitting intentions. We believe that this makes intuitive sense: Usually, risks can be reliably mitigated by a human only if
there is an action available to them that averts the risk (\emph{causal power}),
they can actually perform that action (\emph{self-control}), 
they know that the risk is imminent and which action could mitigate it (\emph{epistemic access}), and
they also want to mitigate it (\emph{fitting intentions}). 
This intuition can be put more precisely, as is done in the remainder of this subsection. Additionally, \autoref{fig:effFlowDiagram} displays the four \conditions of effectiveness.

\subsubsection{Causal power} 

To be effective, the human must be able to make a relevant change.
More precisely: 

\begin{definition}[Causal power]\label{def:causalpower}
    \emph{The agent has the power to establish a sufficient causal connection to the relevant aspect of the world.}
\end{definition}

\noindent Applied to the context of human oversight, this means that \emph{the oversight person has the power to establish a causal connection to parts of the system or its effects, especially the risk}.
For instance, this may include the ability to influence the system, including its components, operation, or outputs. 
Which kind of causal power is appropriate and necessary depends on the context at hand. For example, a factory robot assisting a human in assembly might need a stop button or a manual control mechanism. For an AI tool that assesses university applications and makes recommendations for the admission committee it might be enough if the oversight person can overwrite or disregard the system's output and then forward possible issues to the system provider.

In any case, it is not enough if the oversight person is a mere observer of the system. Without proper means to interfere with it or at least with its effects, they could only watch the manifestation of a risk without being able to mitigate it. Also, it is not sufficient if the human only has `pseudo-options', i.e., if they can only \emph{seemingly} influence the system 
but not really make a relevant difference.\footnote{Pseudo-options, though, are not to be confused with merely unattractive but real options, such as 
pushing a stop button for fun and thereby halting an entire production line for no reason.
} Instances of such scenarios are malfunctioning stop buttons, or environments in which the human's actions are by default not considered. 

\subsubsection{Epistemic access} 

To be effective, the oversight person must know what to do and when to do it. More precisely:

\begin{definition}[Epistemic access]\label{def:epistemicaccess} 
    \emph{The agent has sufficient knowledge of their decision situation (also cf.\ \cite{Shaffer2012}).} 
\end{definition}
\noindent Applied to the context of human oversight, this means that \emph{the oversight person has sufficient knowledge about the risk and how to mitigate it, especially knowing a risk mitigation action}.
An oversight person will, for instance, need to have a sufficient understanding of what the system is doing, in which state it is, which possible risks and benefits there are, which means of influencing the system exist, which effects will result from possible interventions, and which of these effects are most desirable. In some cases, this might include knowledge of normative, social, or cultural domains, without which it would be infeasible for the oversight person to assess certain risks correctly. They do not need to have perfect knowledge, though. They might be unsure or even ignorant about some aspects of their decision situation -- as long as they know enough to achieve their relevant objectives, they have enough epistemic access \cite{Speith2023}. Notably, we assume that the oversight person has certain basic knowledge, such as what a stop button is or how to operate a computer on a consumer-level. One measure to improve the epistemic access of the oversight person could arguably be to employ suitable explainability and transparency measures \cite{DBLP:conf/re/SterzBLH21, biewer2023software, DBLP:journals/chb/SchlickerLOBKW21, doppelbaum2023, Jiang2022, Onitiu2023, deck2024mapping}.

\subsubsection{Self-control} 
To be effective, the oversight person must be in charge of their own doing. 
More precisely:

\begin{definition}[Self-control]\label{def:selfcontrol}
    \emph{The agent can decide for any path of action in their decision situation and, if they do so, follow through with it.}\footnote{In the sense of \textit{guidance control} in \cite{Fischer1998-FISRAC-3}.} 
\end{definition}
\noindent Applied to the context of human oversight, this means that \emph{if there is a risk-mitigating path of action, the human can decide for that path, and if they do so, they actually perform the corresponding action(s)}.
For this, the oversight person needs to be in their right state of mind and free in their actions in a relevant sense. They also have to be in a position to retain attention and purposefully target their actions during the oversight. If, for example, the job of overseeing a fleet of autonomous vehicles is so uneventful and dull that a normal person cannot help but let their mind wander off from their task, the oversight would ultimately become ineffective. Only if oversight persons have self-control, they can purposefully address risks. Other reasons why an oversight person could lack self-control would be when they are severely sleep-deprived, under the influence of substances, or suffering from an active seizure.

An oversight person who lacks self-control might also simultaneously lack sufficient causal power or epistemic access (e.g., in case of a seizure). 
Nevertheless, self-control is distinct from the other two \conditions, as an oversight person can have self-control without having causal power or epistemic access. In fact, self-control may, in many situations, be an antecedent of causal power and epistemic access -- but not vice versa.

Sometimes, oversight persons will decide in split-seconds or will instinctively perform a certain move, and we typically regard this as self-controlled. For example, an oversight person of a factory robot who realizes that it is just milliseconds away from severely injuring a human worker might reflexively slam the stop button. 
We consider this to be a real decision, even if made in a split-second; and since the oversight person followed suit with their decision, they are to be considered self-controlled. Therefore, even oversight persons deciding instinctively can be effective according to the upcoming \Cref{def:effectiveness}.\footnote{\label{fn:splitsecond}Alternatively, one might say there was not a true decision made in the example, but rather an instinctive jolt for the stop-button. In this case, the oversight person would not count as self-controlled according to \Cref{def:selfcontrol} (and therefore also not as effective according to \Cref{def:effectiveness}). If one wishes to allow this interpretation of the situation above and simultaneously allow that the human in it was exercising effective oversight, then the definition of self-control might be adapted to say \enquote{could} instead of \enquote{can}. In this case, even certain instinctive movements of the above kind fall under the definition. However, since we want to avoid a discussion of the conceptual idealization introduced by the \enquote{could}, we will stick to \Cref{def:selfcontrol} as is.}

\subsubsection{Fitting intentions} 

To be effective, the human must also want to do their job properly.
More precisely:
\begin{definition}[Fitting intentions]\label{def:intentions}
    \emph{The agent has intentions that are fitting for their role (or some other relevant standard).} 
\end{definition}
\noindent Applied to the context of human oversight, this means that \emph{the  oversight person \emph{pro tanto} intends to mitigate risks, i.e., they intend to mitigate any risks while taking other relevant factors, such as the interests of the system's users and other stakeholders, into consideration in a suitable way.} 
If the oversight person were unwilling to make the efforts necessary to mitigate a risk, they would not be effective in their oversight -- even if they had causal power, epistemic access, and self-control. This may, for example, be the case if the oversight person lacks the motivation to do their job properly or has other conflicting interests. Even worse, if an oversight person had ill intent, they could use their position to foster risks instead of mitigating them. For example, a racist overseeing a university admission system could deliberately overwrite outputs where people of color are ranked highly. 
In this case, their oversight should be considered ineffective in the relevant respects. 

\subsubsection{Effectiveness} \label{sec:vagueness}

The above yields the following conceptualization of effectiveness in human oversight:

\begin{definition}[Effectiveness]\label{def:effectiveness}
An oversight person is effective in their human oversight if and only if they have \emph{causal power}, \emph{epistemic access}, \emph{self-control}, and \emph{fitting intentions} in the above senses.
\end{definition}

\noindent Conceptualized this way, there still is considerable vagueness and generality to \enquote{effective human oversight}, for example, due to the term \enquote{sufficient} in Definitions \ref{def:causalpower} and \ref{def:epistemicaccess}, or the \enquote{fitting} in \Cref{def:intentions}. 
This intentional vagueness is not a drawback but rather an essential feature. Not every oversight context requires the same level of epistemic access or causal power and, hence, the same exercise of oversight might count as effective in one context but not in another. Accordingly, our conceptualization of effectiveness allows for adaptability and versatility, acknowledging the complexity and diversity of situations where human oversight is applied.

Moreover, \enquote{effectiveness} is not a binary concept but a matter of degrees: one exercise of oversight might be \emph{more effective} than another. Therefore, considering effectiveness in human oversight as a multifaceted continuum suitably encapsulates its nuanced nature. In line with the literature \cite{green-flaws} and legislation \cite{ai-act-EC-2021}, we nevertheless allow the use of \enquote{effective} as a binary concept.  
Saying that an oversight person is effective is to mean that they are \emph{effective enough} for the concrete context they are operating in. In essence, this is also what \Cref{def:effectiveness} says. While it is an open and important question of how much effectiveness is enough for a given context, we will not discuss this issue as it would go far beyond the scope of this paper.

\subsection{Further considerations}

\subsubsection{Competence}
Another aspect of an oversight person that could potentially come to mind is their \emph{competence}. Intuitively, a competent oversight person will be better at being effective. While this is true, there is good reason why competence is not part of our definition of effectiveness: roughly speaking, someone is competent in performing a task if they know what to do to accomplish the task and are actually able to do it in a relevant sense. So, competence is tightly related to the \conditions of effectiveness, even though competence is not itself part of the definition of effectiveness. A competent oversight person will be more likely to establish and retain sufficient epistemic access, self-control, and to use their causal powers in the right way. Without the necessary competence, an oversight person will arguably be unlikely to be effective, if they even can be effective at all.

\subsubsection{Responsibility}
\label{sec:responsibility}

Our conceptualization of effectiveness in human oversight already accounts for responsibility while also going beyond that. 
Taking inspiration 
from a definition of moral responsibility in computing in \cite{sep-computing-responsibility} and being consistent with the general idea of other sources \cite{sep-moral-responsibility, Fischer1998-FISRAC-3} we can define responsibility as follows: 
    An agent is (morally) responsible for some aspect of the world if and only if
    \begin{enumerate*}
        \item the agent has the power to establish a sufficient causal connection to this aspect of the world,
        \item the agent has sufficient knowledge of their decision situation, and
        \item the agent can decide for any path of action in their decision situation and, if they do so, follow through with it.
    \end{enumerate*}
In other words: an oversight person is morally responsible for risk mitigation (or any other part of their job) if and only if they have
   \begin{enumerate*}
        \item causal power,
        \item epistemic access, and
        \item self-control in that respect.
    \end{enumerate*}  
This can be easily translated to cover past events: an oversight person was responsible for, say, mitigating risk if and only if they had sufficient causal power, epistemic access, and self-control in that respect. These insights lead to the more concise formula that
\par
\vspace{0.5em}
\begin{center}
    An oversight person is effective if and only if they are morally responsible and have fitting intentions.\footnote{If \Cref{def:selfcontrol} were adapted to say \enquote{could}, instead of \enquote{can} as suggested in \cref{fn:splitsecond}, this slogan would need to be altered to exclude cases of split-second, reflexive movements.}
\end{center}
\vspace{0.5em}

\noindent This account also allows for shared responsibility between multiple agents, which is desirable in the context of high-risk AI systems. If, for example, both an oversight person and the deployer had causal power, epistemic access, and self-control with regard to a certain risk, they would both be responsible. For reasons of simplicity, the definition does not allow for group responsibility but can be adapted to do so.\footnote{
Our conceptualization of responsibility should not be considered a blueprint for the legal domain. While it may share similarities with some (though not all) existing legal definitions of responsibility, caution is warranted against equating the two, since legal definitions of responsibility vary significantly across diverse areas of law, legal systems, and legal cultures.}  

\section{Facilitators and Inhibitors of Effective Human Oversight}\label{sec:facilitators}

Beyond providing a general conceptualization, our proposed definition of effectiveness in human oversight has a practically useful dimension. Specifically, its \conditions can stimulate ideas about possible facilitators and inhibitors of effectiveness. Facilitators are expected to promote effective oversight, while inhibitors are expected to hinder it. We organize them into three categories: technical design features, individual factors of oversight persons, and environmental circumstances (see, e.g., \cite{Parker_2020} for a similar scheme).  The purpose of the discussion in this section is to provide readers with a more concrete understanding of possible factors regarding effective human oversight in accordance with our suggestions. The layout of \Cref{tbl:mappingFacilitatorsResponsibility} could serve as a template for thinking through possible facilitators and inhibitors, which can be useful to designers and deployers of AI-based systems. The following overview of facilitators and inhibitors is not meant to be exhaustive, and some of them may partially overlap each other. 
Also, they generally do not refer to a specific use case and some may not be applicable to every situation. 

\begin{table*}
    \small
    
    \noindent \begin{tabular}{r c c c c c | c c c c c c | c c c c c } 
         &
        \rot{\hyperlink{fac:interventionOptions}{intervention options}} &
        \rot{\hyperlink{fac:systemAdaptability}{system adaptability}} &
        \rot{\hyperlink{fac:systemUnderstandability}{system understandability}} &      
        \rot{\hyperlink{fac:interpretability}{interpretability of in- and outputs}} &
        \rot{\hyperlink{fac:preselection}{preselection of outputs to review}} &       
        \rot{\hyperlink{fac:overseerTraining}{training of the oversight person}} &        
        \rot{\hyperlink{fac:domainExpertise}{domain expertise}} &
        \rot{\hyperlink{fac:conscientiousness}{conscientiousness}} &        
        \rot{\hyperlink{fac:exhaustion}{exhaustion}} &
        \rot{\hyperlink{fac:motivation}{motivation}} &
        \rot{\hyperlink{fac:automationBias}{automation bias}} &     
        \rot{\hyperlink{fac:jobDesign}{adequate job design}} &      
        \rot{\hyperlink{fac:roleConflicts}{role conflicts}} &
        \rot{\hyperlink{fac:independentThinking}{independent thinking}} &
        \rot{\hyperlink{fac:accountability}{accountability}} &
        \rot{\hyperlink{fac:timePressure}{time pressure}}\\  \hline
        
         & \multicolumn{5}{c|}{technical design} & \multicolumn{6}{c|}{individual factors} & \multicolumn{5}{c}{environment} \\\hline
         
        causal power         
        & \tablebullet
        & \tablebullet 
        &  
        &  
        &  
        & \tablebullet 
        & \tablebullet 
        &  
        &  
        &  
        &  
        &  
        &  
        &  
        &  
        & \tablebulletNegative 
        \\ 
        
        epistemic access 
        &  
        & \tablebullet 
        & \tablebullet 
        & \tablebullet 
        & \tablebullet 
        & \tablebullet 
        & \tablebullet 
        & \tablebullet 
        & \tablebulletNegative 
        & \tablebullet 
        & \tablebulletNegative 
        & \tablebullet 
        &  
        & \tablebullet 
        & \tablebullet 
        & \tablebulletNegative 
        \\ 
        
        self-control 
        &  
        &  
        &  
        &  
        &  
        & \tablebullet 
        &  
        & \tablebullet 
        & \tablebulletNegative 
        & \tablebullet 
        & \tablebulletNegative 
        & \tablebullet 
        &  
        &  
        & \tablebullet 
        &  
        \\ \hline
        
        fitting intentions 
        &  
        &  
        &  
        &  
        &  
        & \tablebullet 
        &  
        & \tablebullet 
        & \tablebulletNegative 
        & \tablebullet 
        & \tablebulletNegative 
        & \tablebullet 
        & \tablebulletNegative 
        &  
        & \tablebullet/\tablebulletNegative 
        &  
        \\ \hline \\
        
       \end{tabular}
       
       \caption{Overview of how exemplary facilitators (\tablebullet) and inhibitors (\tablebulletNegative) usually contribute to the effectiveness of human oversight.}
       \label{tbl:mappingFacilitatorsResponsibility}
\end{table*}
 
\subsection{Technical design features}
Under \enquote{technical design features}, we subsume facilitators and inhibitors whose influence on effective human oversight results from the design of the system. This list could also include aspects such as interface and interaction design \cite{DBLP:conf/chi/Norman83}, runtime monitoring systems~\cite{rtlolacavindustrial}, fault tolerance \cite{koren2020fault}, or automated anomaly detection systems~\cite{DBLP:journals/access/NassifAND21}.

\hypertarget{fac:interventionOptions}{\textbf{Intervention options}}. For an oversight person to have causal power over the system, there must be options that allow the oversight person to intervene with, control, overwrite, or undo system decisions and actions. For example, an intervention option might be a stop button that allows oversight persons to stop system operation at any time. Other examples include options to take over manual control or to initiate emergency protocols that result in fail-safe modes.

\hypertarget{fac:systemAdaptability}{\textbf{System adaptability}}. It can be beneficial to oversight if oversight persons can adapt the system's processes or its interface~\cite{Evangelista_Belo_2022,Sauer_2017}. For example, the oversight person could reduce the speed at which the system produces its outputs or adapt the system's interface to display additional information. This increases the oversight person's epistemic access if the adaptations help him to acquire and process the information better.

\hypertarget{fac:systemUnderstandability}{\textbf{System understandability}}. Under \enquote{system understandability} we subsume design aspects associated with the area of explainable AI~\cite{Langer_2021}. For example, this includes the use of inherently transparent algorithms, the use of approaches to explain system outputs, the use of what-if analyses, or the provisioning of log files to enhance traceability of system processes~\cite{arrieta2020explainable, Speith_2022}. All options for enhancing system understandability aim at improving epistemic access~\cite{Langer_2021,hoffman2018metrics}.

\hypertarget{fac:interpretability}{\textbf{Interpretability of inputs and outputs}}. By \enquote{interpretability}, we mean a representation of inputs or outputs that is appropriate for human understanding~\cite{Rudin_2019}. This could include, for example, the provision of information about what input features mean, or visualizing input and output features in a way that makes them easier to grasp. Interpretability thus aims at improving epistemic access.

\hypertarget{fac:preselection}{\textbf{Preselection of outputs to review}}. There may be cases where oversight tasks require a mechanism for selecting system outputs that are to be reviewed by the oversight person. 
By defining boundary conditions or by using oversight support tools, oversight persons could gain better epistemic access when only having to review a subset of outputs. For example, in a hiring context, such a tool could be used to identify applicants for whom the AI-based evaluation was obviously fair and free of bias, leaving considerably fewer cases to the oversight person~\cite{biewer2023software}. 

\subsection{Individual factors of oversight persons} 

Under \enquote{individual factors of oversight persons} we subsume traits and states of oversight persons as well as interventions \emph{on} oversight persons that could facilitate or inhibit effective oversight. This discussion may cover aspects like vigilance~\cite{Warm_2008}, cognitive abilities \cite{spearman1904general, Warne_2018}, humans' propensity to trust in automation \cite{hoff2015trust}, and additional cognitive biases \cite{Roese_2012} and heuristics \cite{Tversky_1973}.

\hypertarget{fac:overseerTraining}{\textbf{Overseer Training}}. Training to prepare for or enhance the role of the oversight person will be crucial. For instance, training in which people are confronted with system errors or erroneous automated decisions can help them to become better at identifying these errors when later acting on the real system~\cite{Bahner_2008}. Overseer training could, in principle, target all of the \conditions of effective human oversight. Such training could aim to enable oversight persons to maintain epistemic access, to recognize earlier when they are losing self-control (e.g., become inattentive \cite{Perels_2005}), to perform certain maneuvers that increase their causal power, or to know what the job requires them to do and thereby improve the fit of their intentions.

\hypertarget{fac:domainExpertise}{\textbf{Domain expertise}}. Overseers may require a certain degree of expertise in the task performed by the system they are supposed to oversee. For instance, physicians with substantial expertise in diagnosing a disease may have better epistemic access (e.g., they will know better when a system is making incorrect diagnoses) than physicians with little expertise \cite{Gaube_2023}. Domain expertise might sometimes also enable more causal power.  

\hypertarget{fac:conscientiousness}{\textbf{Conscientiousness}}. Conscientiousness is a character trait of people who are self-disciplined, orderly, goal-directed, who make and follow plans, and who adhere to social norms \cite{Costa_1991}. Highly conscientious oversight persons will arguably be more likely to maintain self-control, sufficient epistemic access, and fitting intentions. 

\hypertarget{fac:exhaustion}{\textbf{Exhaustion}}. Exhaustion is one of many psychological strain symptoms~\cite{Sonnentag_2023}. It is a state where an oversight person feels tired and worn out~\cite{maslach1997maslach} and thus may not be good at maintaining self-control and sufficient epistemic access. 

\hypertarget{fac:motivation}{\textbf{Motivation}}.\label{sec:motivation} Motivation refers to internal and external factors that drive and direct an individual's behavior towards a goal \cite{Ryan_2000}. Motivation leads people to initiate, intensify, and persist with goal-directed behavior. Motivation is often influenced by individual factors, environmental factors, and the perceived significance of the desired outcomes. In the context of human oversight, motivation is crucial to initiate and maintain fitting intentions and self-control for the goal of trying to mitigate risk. It is also important for persisting in maintaining sufficient epistemic access.

\hypertarget{fac:automationBias}{\textbf{Automation bias}}.\label{sec:automationbias} Automation bias is usually considered a cognitive bias where people use the outcome of an automated decision aid \enquote{as a heuristic replacement for vigilant information seeking and processing} (\cite{Mosier_1996}, p. 205). Automation bias becomes especially likely when facing a system that works reliably and with high performance because, in such cases, it is rational for people to divert their attention to other tasks \cite{Parasuraman_2010}.
This, however, will contribute to insufficient epistemic access. If there are indeed no negative consequences for some time, this may lead to what Parasuraman and Manzey \cite{Parasuraman_2010} called \enquote{learned carelessness}. This phenomenon could be interpreted as a problem of self-control or as an issue that leads to intentions that no longer fit the duties of an oversight person. 

\subsection{Environmental circumstances}

Under \enquote{environmental circumstances}, we subsume factors that lie outside the technical system or the oversight person, yet they can influence the effectiveness of oversight. This discussion could also include aspects such as stakes of the situation~\cite{Langer_2021a}, single vs. multitasking environments~\cite{Lyons_2018}, or whether oversight is organized as a team task \cite{Gao_2014,Ulfert_2023}.

\hypertarget{fac:jobDesign}{\textbf{Adequate job design}}. The job of an oversight person needs to be well designed \cite{Parker_2020}, which means that the job is, e.g., motivating, satisfying, and not overly demanding. For example, the job characteristics model from work psychology~\cite{Hackman_1976} highlights that certain job characteristics promote motivation and satisfaction with the job, namely skill variety, task identity, perceived task significance, autonomy, and feedback from the job.\footnote{See also other models from psychology that inspire job design to make the work more motivating and satisfying (e.g.,~\cite{Humphrey_2007}), or less demanding (e.g. the job-demands-resources model~\cite{Demerouti_2001}).} 
The design of the job is important for promoting self-control, epistemic access, and fitting intentions, for example via motivation (cf.\ \Cref{sec:motivation}). 
Also, ill-designed jobs can lead to counterproductive work behavior, i.e., behavior at work that is inconsistent with or even intentionally disrespecting job duties (e.g., not even trying to maintain sufficient epistemic access, or intentionally overwriting outputs to feel \enquote{needed}~\cite{Strich_2021}).  

\hypertarget{fac:roleConflicts}{\textbf{Role conflicts}}. Role conflicts could arise if the oversight person is at the same time a decision-maker who uses the AI-based system to support their decisions \cite{King_1990}. Consider, for example, a hiring scenario where an individual juggles both the role of an oversight person tasked with mitigating unfairness and the role of an HR manager who is responsible for inviting only the best candidates to a job interview. A conflict arises when the only method to address unfairness involves inviting additional, apparently less suitable candidates, conflicting with the HR manager's goal of selecting the best candidates only. This misalignment of duties may give rise to intentions that are unfitting for an oversight person.

\hypertarget{fac:independentThinking}{\textbf{Independent thinking}}. Processing a situation under the impression of a system's output can affect an oversight person's judgment of such an output \cite{Furnham_2011}. For example, seeing a recidivism risk score for a defendant in court, in addition to the information about the defendant's case, may affect how people judge this case. Seeing a \enquote{high anchor} (i.e., the system's prediction of a high risk for recidivism) will make it more likely that the judge will decide that the defendant has a high risk for recidivism. 
Such anchors make oversight persons selectively aware of anchor-consistent information \cite{Englich_2006} which undermines epistemic access.

In the area of AI-supported decision-making, there are several ideas of how to possibly prevent such effects. For example, although this leads to less acceptance of the system, it seems to partly counter such anchoring effects if people are forced to think about a given case themselves for some time before making a final decision \cite{Bu_inca_2021}. 
Alternatively, 
oversight persons may only receive the AI-based output after they have made up their own mind about a specific case because then they would not initially be affected by the AI-based output and may be in a better position to judge whether the AI-based output is appropriate \cite{green-flaws, Zerilli_2019}.

\hypertarget{fac:accountability}{\textbf{Accountability}}. Here we refer to  accountability as “an obligation to explain and to justify [one’s] conduct” \cite[p.447]{Bovens_2007}. If oversight persons feel accountable for their actions 
this may influence their attentiveness or may lead to them  feeling more stressed \cite{Hall_2015}. Thus, a feeling of accountability could promote efforts of oversight persons to get themselves in a sufficient epistemic position. It can also be hypothesized that a felt obligation to explain oneself will motivate to maintain high self-control in order to keep up with the expectations of the job and that it will incentivize one to uphold intentions that fit one's duties. However, it could also lead to unintended consequences such as leading to unfit intentions: if oversight persons believe that they would have no good explanation for overriding a system output, they may not do so in cases where they know that they have to file a report after overwriting system outputs.

\hypertarget{fac:timePressure}{\textbf{Time pressure}}. Significant time pressure may undermine various \conditions for effectiveness \cite{Rieger_2022}. For example, if an oversight person only has a few seconds to react to warnings by the system, this could hinder sufficient epistemic access. In extreme cases, time pressure could also undermine an oversight persons' causal power. For instance, if the option to overwrite an output is only available for split seconds, the oversight person may not be able to press the button.

\section{Application to the AI Act Proposal}\label{sec:AIAct}

Finally, our analysis turns to effective human oversight as outlined in the AI Act~\cite{ai-act-ParliamentCOR}\footnote{At the time of writing, the trilogue negotiations between the European Commission, the Parliament, and the Council have been concluded and a finalized version has been approved by the European Parliament. However, the final version of the AI Act has not yet been approved by the Council of the EU.}. According to Article 14(1) of the proposal, what the AI Act defines as high-risk AI systems shall be designed and developed in such a way that they can be effectively overseen during their use. As stated in Article 14(2), this oversight shall aim to prevent or minimize risks to health, safety, or fundamental rights. Risk mitigation is therefore defined as a key objective of human oversight measures that should contribute to the overall trustworthiness of high-risk systems \cite{enqvist2023human}. Consequently, the effectiveness of human oversight, as outlined in Article 14(1), serves as a foundational principle that requires oversight measures to be capable of achieving their risk mitigation objective. This principle can be understood as a condition for a minimum level of effectiveness of the oversight measures implemented, comparable to a threshold that must be reached. 

\subsection{Conditions for effectiveness in our proposal and the AI Act}

Article 14(3) obliges the provider to enable its deployer to perform effective oversight by identifying \enquote{appropriate measures}.\footnote{It should be noted that the Commission proposal \cite{ai-act-EC-2021} referred to the deployer as \enquote{user}. The European Parliament \cite{ai-act-Parliament} rightly proposed to use the term \enquote{deployer} instead to avoid any confusion with the end user of the AI system. This proposal found its way into the version approved by the Parliament \cite{ai-act-ParliamentCOR}. We therefore use the term \enquote{deployer}.} 
Article 14(4) contains certain aspects of what the proposed AI Act envisages as indicators of the effectiveness of these measures. These require enabling the individuals to whom human oversight is assigned to do the following, as appropriate and proportionate to the circumstances:

\begin{quote}\small

(a) to properly understand the relevant capacities and limitations of the high-risk AI system and be able to duly monitor its operation, including in view of detecting and addressing anomalies, dysfunctions and unexpected performance; \\

(b) to remain aware of the possible tendency of automatically relying or over-relying on the output produced by a high-risk AI system (automation bias), in particular for high-risk AI systems used to provide information or recommendations for decisions
to be taken by natural persons; \\

(c) to correctly interpret the high-risk AI system’s output, taking into account, for example, the interpretation tools and methods available; \\

(d) to decide, in any particular situation, not to use the high-risk AI system or to otherwise disregard, override or reverse the output of the high-risk AI system; \\

(e) to intervene in the operation of the high-risk AI system or interrupt the system through a ‘stop’ button or a similar procedure that allows the system to come to a halt in a safe state. \\

{\small
(quoted from Article 14(4) of the amendment to the version approved by the European Parliament \cite{ai-act-ParliamentCOR})\\}

\end{quote}

\noindent These measures reflect our \conditions of effectiveness. 
Requirements (d) and (e) underscore the EU's recognition of the importance of oversight persons being able to influence the AI system, 
either by modifying the system's output or by stopping its operation. These requirements
manifest the need for both causal power and self-control, ensuring that individuals can influence the decisions of the AI system, or decide not to use the system's output or to modify it. Conditions (a) and (c) relate to the understanding of the system, as they require the provider to enable the oversight person to understand  the limitations of the system (though not necessarily all details of the system), to monitor it effectively, and to interpret its results correctly. This is consistent with the condition of epistemic access, which enables the oversight person to make informed decisions. Awareness of automation bias, as described in (b), is intertwined with epistemic access and self-control, as described above in \Cref{sec:automationbias}. 
Furthermore, Article 26(2) \cite{ai-act-ParliamentCOR} emphasizes the need for the necessary competence, training, authority, and support in oversight functions, hence requiring epistemic access and causal power.\footnote{This requirement was mentioned only in the non-binding recital (48) of the EU Commission's proposal \cite{ai-act-EC-2021}, but was ultimately stipulated more directly in Article 29 of the proposals by the EU Council \cite{ai-act-Council}, the Parliament \cite{ai-act-Parliament} and Article 26 of the approved version \cite{ai-act-ParliamentFINAL}.}
The criterion of fitting intentions of the agents involved has little direct equivalent in the legal text, which is not focused on subjective intentions but on actions. The parties' underlying intentions and motivations are not directly relevant. However, one might consider that the automation bias targeted in Article 14(4) (b) may interfere with the fittingness of the oversight person's intentions (cf.\ \Cref{sec:automationbias}). Furthermore, the significant penalties according to Article 99 \cite{ai-act-ParliamentCOR} strongly incentivize norm-compliant behavior.

\subsection{Further analysis of the EU legislation's approach}

\paragraph{Strongly varying levels of abstraction} It is striking that the measures appearing in Article 14(4) are of varying degrees of abstraction: While some measures contain specific requirements for the design of the AI system, other measures are less specific. For example, the requirement of a \enquote{stop} button in (e) is highly concrete. In contrast, the requirement in (a) to create an understanding of the limitations of the system is vague and leaves a lot of room for interpretation.  Overall, the requirements for the various parties -- providers of the AI system, deployers, and the person entrusted with oversight -- could have been made clearer in the structure of the AI Act. Thus, Article 14(4) appears to be a loose collection of items that were deemed useful in the legislative process without much structural consistency. The lack of regulatory clarity could hinder the practical applicability of Article 14 and, thus, the introduction of truly effective human oversight. 

\paragraph{Significance of Psychological Biases} 
Measures in accordance with (b), which are intended to maintain an awareness of the automation bias, seem out of place: it introduces a concrete psychological aspect that sticks out from the more technical and operational aspects outlined in the other provisions, especially considering that automation bias is just one of many relevant biases (e.g., hindsight bias, confidence bias \cite{Hilbert_2012,Roese_2012}), heuristics (e.g., availability heuristic \cite{Tversky_1973}), and other psychological phenomena (e.g., exhaustion, motivation \cite{Ryan_2000,Sonnentag_2023}) that would need to be accounted for.
\paragraph{Legal uncertainties} Not every measure in Article 14(4) will have to be fully implemented in every high-risk system. This already follows from the restriction that the measures should be implemented
\enquote{as appropriate and proportionate to the circumstances}.
As a result, the specific measures taken must be assessed on a case-by-case basis. Not only \emph{how} these requirements apply, but also \emph{which} requirements are applicable depends heavily on the individual case and the design of the system. Although AI Act's approach provides flexibility that accommodates the sometimes beneficial vagueness of the effectiveness of human oversight also mentioned in \Cref{sec:vagueness}, it may also be associated with greater legal uncertainty. 
The legislation will eventually be complemented by technical standards developed by European standardization organizations to at least partially remove this legal uncertainty and to facilitate the implementation of requirements for high-risk systems, including human oversight \cite{VealeBorgesius}. However, this standardization process is still in its early stages and is limited to a few general guidelines \cite{laux2023}. Moreover, the question arises as to what requirements these standards must meet to satisfy the
effectiveness requirement of Article 14(1). While Article 14(4) offers some guidance, its vagueness and varying levels of abstraction are insufficient for determining true effectiveness in specific cases. Thereby, the standardization organisations risk developing technical standards that may not withstand legal scrutiny. So, the EU's reliance on the AI Act's effective implementation through practical technical standards poses a potential threat to the Regulation's requirement for human oversight.

\paragraph{The practical need for a definition} 
Also, the technical standards created in this process only have a presumptive effect for providers and deployers of AI systems that are subject to the AI Act. In the event of a legal dispute, the courts will therefore have to examine whether the specific design of the human oversight measures meets the overarching requirement of effectiveness in Article 14(1). As mentioned above, this will depend on the circumstances of the individual case. To ensure uniform application of the AI Act by lower courts, a comprehensible definition of effective human oversight is essential. In this respect, the definition of effectiveness will be one of the key tasks of the deciding courts in the coming years, leaving these courts also in need of further clarification on the conditions of effectiveness.

\subsection{Utilizing our definition of effectiveness}

Our objective-first approach to effectiveness can be used to integrate the piecemeal list of different requirements listed in the AI Act into a coherent system. This is advantageous for a legally sound application of the AI Act, as it helps to structure the requirements for and responsibilities of the different parties involved. One of the benefits of the breadth and structure of our conceptualization is that it simplifies the assessment process for regulators, making it easier to evaluate different AI systems against a structured set of principles that are framed by the four \conditions. The three domains of facilitators and inhibitors can aid in giving further structure for certain use cases. This streamlining may also decrease the legal uncertainty in Article 14, fostering a more consistent and predictable environment for AI providers and deployers alike, while providing a practical definition of effectiveness for deciding courts. This could also help uncover further shortcomings in Article 14(4): For instance, while understanding the system's output (c) can enhance epistemic access, understanding of the system's \emph{input} can be equally crucial. Our conditions for effectiveness, when applied with expertise, make this need evident; whereas this need can be challenging to identify when relying solely on Article 14(4), which does not mention the system's input at all.

Consequently, the points listed in \Cref{tbl:mappingFacilitatorsResponsibility} can be understood as examples of how an assessment of the effectiveness of measures could be carried out. Defining effectiveness using our four conditions provides a practical framework that highlights the key aspects of effective oversight. Our approach is, therefore, well suited to assist standard-setting organizations and to clarify the rather broad requirements of Article 14(4). In addition, our proposed framework provides guidance to practitioners who do not wish to rely solely on technical standards. Finally, our approach can assist courts in the future by providing a framework for defining and assessing the effectiveness of human oversight.

\section{Limitations and Future Work}
\label{sec:limitations}
As discussed above, we see advantages brought about by our conceptualization of effective human oversight. Nevertheless, we also want to point out three limitations as well as future points to address. 
\emph{First}, our conceptualization is designed for broad applicability across various AI-based systems, and thus shares a common criticism with the AI Act, namely that of residing on a high level of abstraction and that of vagueness, as discussed in \Cref{sec:vagueness}. Dealing with these issues, especially in complex scenarios, requires (a) involving individuals with extensive expertise in the task at hand and (b) fostering collaboration in multidisciplinary teams to assess the impact on human oversight practice. 
\emph{Second}, when we came up with the facilitators and inhibitors spelled out in \Cref{sec:facilitators}, we realized that our conceptualization is helpful for thinking through conditions that could promote or hinder effective oversight. However, for some of the facilitators (or inhibitors), the question of which of the four conditions they contribute to (or hinder) depends on the concrete context. Moreover, for certain cases, the mapping from facilitators and inhibitors to \conditions needs considerable empirical research, outside the scope of this conceptual paper. 
\emph{Third}, we did not discuss the concept of human oversight itself in much detail, but focused only \emph{effectiveness} of human oversight. So, while our paper is helpful in determining whether an instance of human oversight is effective, it is less helpful in assessing whether something is human oversight. However, our definition of effectiveness is broad enough to be applied to many roles along the AI pipeline, not only to oversight persons. Some more conceptual discussion of human oversight itself is, for instance, provided by \cite{laux2023}. 

Future efforts will aim to enhance the practical applicability of our framework. We plan to develop a comprehensive practical framework, help to establish best practices, and create tools for implementing and auditing effective human oversight. These developments can benefit both policymakers and practitioners. This endeavor will require ongoing interdisciplinary collaboration, such as conducting empirical studies to validate and refine oversight mechanisms and designing holistic requirements for high-risk systems, oversight personnel, and their environments. Additionally, efforts should include creating educational programs and training for oversight persons, as well as engaging with regulatory bodies to ensure the ethical deployment of AI systems.

\section{Conclusion}
\label{sec:conclusion}

This paper has proposed an approach to effective human oversight -- encompassing causal power, epistemic access, self-control, and fitting intentions --  to remedy the lack of clarity surrounding the concept of effectiveness. This is a joint contribution of researchers in  psychology, law, philosophy, and computer science. We argued that, in essence, a morally responsible oversight person with fitting intentions is generally suitable for mitigating risks associated with high-risk AI systems.
We have identified facilitators and inhibitors of effectiveness in three categories, namely the technical design of the system, individual characteristics of oversight persons, and the environmental circumstances in which oversight occurs. Thereby, we have provided inspiration for steps towards a successful implementation of effective human oversight.
We also discussed the extent to which our understanding of effective human oversight aligns with the European Union's upcoming AI Act, in particular Article 14 on human oversight. Our analysis suggests that the AI Act could benefit from incorporating more nuanced and structured conditions, such as those proposed in this paper.

In closing, we want to advocate for a balanced approach to effective human oversight. Without a concrete idea of what effectiveness is in the context of human oversight, no productive discussion of it is possible. On the one hand, we urge to not throw out the baby with the bathwater by per se dismissing human oversight as a safeguard without the necessary conceptual clarity. On the other hand, caution is warranted as well when emphatically including a notion of effective human oversight in regulations and guidelines without this conceptual clarity. 
Our interdisciplinary approach is intended to aid both advocates and critics of human oversight. For many relevant oversight contexts, the four conditions of effectiveness seem to provide a better picture of when and how human oversight is useful. 
We, therefore, believe that these conditions could form a pivot for future research on human oversight.

\begin{acks}
This work is partially funded by DFG grant 389792660 as part of TRR~248 -- \href{https://perspicuous-computing.science}{CPEC}, by 
VolkswagenStiftung as part of grants AZ 98514, 98513 and 98512 -- \href{https://explainable-intelligent.systems}{EIS}, and by the European Regional Development Fund and the Saarland within the scope of \href{https://certain.dfki.de/}{(To)CERTAIN}.
\end{acks}

\bibliographystyle{ACM-Reference-Format}
\bibliography{main}

\end{document}